\documentstyle[12pt,aasms4]{article}

\def\power#1{\mbox{$\times10^{#1}\ $}}

\newcommand{\lil}{$^{6}$Li }
\newcommand{\li}{$^{7}$Li }
\newcommand{\be}{$^{7}$Be }
\newcommand{\he}{$^{3}$He }
\newcommand{\msun}{M$_\odot$ }
\newcommand{\msyr}{M$_\odot$.yr$^{-1}$ }

\newcommand{\zaa}{A\&A~}
\newcommand{\zapj}{ApJ~}
\newcommand{\zapjs}{ApJS~}

\begin{document}

\title{On the \li and \be synthesis in novae}

\author{Margarita Hernanz}
\affil{Centre d'Estudis Avan\c{c}ats de Blanes (CSIC), Cam\'{\i} de Sta.
B\`arbara, s/n, E-17300 Blanes, SPAIN}
\author{Jordi Jos\'e}
\affil{Departament de F\'{\i}sica i Enginyeria Nuclear (UPC), Avda.
V\'{\i}ctor Balaguer, s/n, E-08800 Vilanova i la Geltr\'u, SPAIN}
\author{Alain Coc}
\affil{Centre de Spectrom\'etrie Nucl\'eaire et de Spectrom\'etrie de
Masse, IN2P3-CNRS, B\^at.104, F-91405 Orsay Campus, FRANCE}
\and
\author{Jordi Isern}
\affil{Centre d'Estudis Avan\c{c}ats de Blanes (CSIC), Cam\'{\i} de Sta.
B\`arbara, s/n, E-17300 Blanes, SPAIN}

\received{}
\accepted{}

\slugcomment{\underline{Submitted to}: \zapj Letters~~~~\underline{Version}:
\today}

\begin{abstract}

The production of \li and \be during the explosive hydrogen--burning
that occurs in nova explosions is computed by means of a hydrodynamic
code able to treat both the accretion and the explosion stages.
Large overproduction factors with respect to solar abundances are
obtained, the exact value depending mainly on the chemical composition
of the envelope. Although the final ejected masses are small, these results
indicate that novae can contribute to the \li enrichment of
the interstellar medium. Furthermore, since \be decays emitting a gamma--ray
(478 KeV), with a half--life of 53.3 days, the synthesis of \li could be
tested during the INTEGRAL mission.

\end{abstract}


\section{Introduction}

The origin of lithium and other light elements is still an unsolved
problem in astrophysics. It is widely accepted that \li isotopes are
produced  during the Big Bang and by "spallation" reactions in the
interstellar medium by galactic
cosmic rays or in flares (see \cite{Ree93} for a recent review).
Standard Big Bang nucleosynthesis underproduces \li with respect to
solar by more than an order of magnitude (see however the recent paper
by \cite{Del95}), whereas spallation reactions by
galactic cosmic rays, produce \li and \lil simultaneously, as well as
$^{9}$Be, $^{10}$B and $^{11}$B.
These two mechanisms are unable to account alone for the
present \li abundance (\li/H $\approx$ 2\power{-9}). Furthermore, they
are unable to produce neither the high isotopic ratio \li/\lil observed in the
solar system (\li/\lil = 12.5 $\pm$ 0.2) nor the $^{11}$B/$^{10}$B one
($^{11}$B/$^{10}$B$\approx$ 4). Recent measurements of the lithium isotopic
ratio in the
interstellar medium (\cite{Lem93}, \cite{Mey93}) yield values similar to
those found in the solar system, indicating that it has remained nearly
constant or even decreased during the last 4.5--5 Gyr. The contribution of
a low energy component of the galactic cosmic rays, confined at the source
or by stellar flares (\cite{Men71}, \cite{Can75}, \cite{Pra93}) can account
for the boron isotopes but \li is still underproduced.
Therefore, an extra stellar source able to produce this  \li without
generating \lil has to be invoked. The interplay of these sources in the
galactic evolution of lithium has been extensively studied (\cite{DAM91},
\cite{Abi95}).

The synthesis of \li by a stellar source requires the formation of \be\,
which transforms into \li by an electron capture, being \be half--life
53.3 days. As \li is very easily destroyed, \be has
to be transported to zones cooler than those where it was formed with a time
scale shorter than its decay time. This {\em beryllium transport} mechanism,
as first suggested by \cite{Cam55}, requires a dynamic situation, like that
encountered in asymptotic giant branch (AGB) stars and novae.
Another possibility is the production of lithium and boron isotopes by
neutrino induced synthesis, during gravitational supernova explosions
(\cite{Woo90} and \cite{Woo95}). The importance of such a mechanism is
still a matter of debate (\cite{Mat95}).

The production of lithium in AGB stars has been extensively studied and these
stars represent the unique observational evidence of an autogenic stellar
origin, since it has been observed in them (\cite{Abi91}, \cite{Abi93}). The
huge abundances of lithium found in some AGB stars are a clear proof that
these stars are currently injecting important quantities of lithium to the
interstellar medium. However it is hard to estimate their total contribution,
since it depends on the estimated number of such stars, that are buried
by their own wind (\cite{Abi93}).

The production of \li in explosive hydrogen burning and, in particular, in
accreting white dwarfs exploding as classical novae, was first studied with
a parametrized one--zone model by \cite{Arn75}. Later on
\cite{Sta78} computed the \li yields by means of a hydrodynamic code.
This code simulated the explosive stage of novae,
without considering the accretion phase, i.e., with an initial envelope already
in place. The conclusion of this work was that, depending on the initial
abundance of \he and on the treatment of convection, \li could be formed in
substantial amounts during explosive hydrogen--burning in novae.
This problem has been revisited by \cite{Bof93}. On the basis of an extended
nuclear reaction network and
of updated nuclear reaction rates, but adopting again a parametrized one-zone
model, they showed that \li could only be produced in significant amounts at
peak densities lower than 10$^{3}$g.cm$^{-3}$, which are lower than those
predicted by hydrodynamic simulations. They argued that the reason of the
discrepancy was the neglect of the $^{8}$B(p,$\gamma)^{9}$C reaction in the
calculations of \cite{Sta78}. However, large overproductions of \be were
found by \cite{novae95} (using the semi--analytical model of \cite{Mac83} to
obtain the temperature and density profiles and a complete reaction network
that included $^{8}$B(p,$\gamma)^{9}$C) showing that the origin of the
different results was not that reaction. In fact, \cite{Bof93} also showed by
means of a two--zone approximation that the efficiency of mixing by convection
is a very critical parameter, and they stressed the need of a detailed
hydrodynamic model to study \li production more accurately.

The purpose of this letter is to compute the synthesis of \li in both carbon--
oxygen (CO) and oxygen--neon--magnesium (ONeMg) novae by
means of an implicit hydrodynamic code, that includes a full reaction network,
able to treat both the hydrostatic accretion phase and the explosion
stage. An estimation of the contribution of novae to galactic enrichment is
made, on the basis of the overproductions and ejected masses obtained. The
importance of the initial chemical composition of the envelope is analyzed.

The detection of \li in novae would confirm our theoretical predicition.
Furthermore, the detection of gamma--ray
emission at 478 KeV, corresponding to the decay of \be to \li (half--life
53.3 days) in the early phases of novae by the future mission INTEGRAL
would also confirm the thermonuclear runaway model for novae and the
nucleosynthesis related to it.

\section{Model and results}

A one--dimensional, lagrangian, implicit hydrodynamic code has been
developed following the techniques described in \cite{Kut72}. The code
has been built
in such a way as to enable the study of both the hydrostatic accretion phase
and the fully hydrodynamic explosion. Detailed nucleosynthesis is obtained
by means of an extended reaction network, including 100 nuclei,
ranging from $^{1}$H to $^{40}$Ca, linked
through an up to date network including more than 370 nuclear reactions (see
\cite{Jos96} and \cite{Job96} for the details). Concerning the reactions
involved in \be synthesis, rates are taken from the \cite{CaF88} compilation,
\cite{Wag69}, \cite{Des89} and \cite{Wie89}. Time--dependent convection
is included in the code, since the hypothesis that the convection time scale
is always shorter than the nuclear time scale, inherent to time--independent
convection, is not always fulfilled. With this method, partial mixing in
the convective region is included.

Complete evolution of the accretion and explosion stages of white dwarfs with
masses ranging from 1 to 1.25 \msun, accreting at a rate of 2\power{-10}\msyr,
with initial luminosity 10$^{-2}$L$_{\odot}$ has been computed.
We assume that the infalling material is of solar composition, but that some
mixing process (diffusion, shear mixing) mixes it with the underlying CO (for
M=1 and 1.15 \msun) or ONeMg (for M=1.15 and 1.25 \msun) core.
This assumption is based on the current prediction that enhanced
CNO (or ONeMg) abundances are required in order to produce a nova outburst
and to explain some observed abundances (see \cite{Liv94} for a recent
review, and \cite{Pri95} and \cite{Pol95},
for recent calculations of CO and ONeMg novae, respectively).
We want to stress that the problem of the initial composition of nova
envelopes is rather complicated and that it is far from being understood in a
self--consistent way. Studies of diffusion during accretion onto CO white
dwarfs have been carried out (\cite{Kov85} and \cite{Ibe92}), but it is not
clear if enough enhancements of heavy elements are obtained in the ejecta.
For the ONeMg white dwarfs, these studies are still lacking. Therefore, a
compromise is to adopt some percentage of mixing with core abundances.
We have adopted a 50\% of mixing by mass with core abundances, as was done
in the work by \cite{Pol95}.

The chain of reactions leading to the formation of \be has been extensively
discussed in \cite{Bof93}.
During hydrogen burning, the formation of \be\ proceeds through
$^3$He($\alpha,\gamma)^7$Be from the initial \he as (p,$\gamma$) reactions
cannot bridge the A=5 gap.
It is destroyed by $^7$Be(p,$\gamma)^8$B, followed by either
$^8$B($\beta^+)2^4$He or $^8$B(p,$\gamma)^9$C($\beta^+$,p)2$^4$He.
However, at high temperature the photodisintegration of $^8$B,
$^8$B($\gamma$,p)$^7$Be, becomes faster than proton capture on $^7$Be.
In these conditions, the effective lifetime of \be\ can become larger than the
time scale of the outburst (\cite{Bof93}). For typical densities
at the base of the envelope at the onset of explosion, this would
happen only above T$_8 \sim 1$. Below this temperature destruction by
$^7$Be(p,$\gamma)^8$B is
efficient.
Other destruction mechanisms are $^7$Be($\alpha,\gamma)^{11}$C and beta decay
to \li.
Radiative alpha capture on \be\ is always slower 
than proton capture as long as photodisintegration of $^8$B is not efficient.
The half-life of \be\ (53.3 days) would allow the formation of \li\ long
after the
outburst when the cooler envelope would prevent the rapid destruction of this
fragile isotope.
Since \be\ is more efficiently destroyed than produced below T$_8\approx$1
and since it originates only from initial \he, it can only be formed during
the outburst if enough \he\ survives the initial phase when the hydrodynamic
time scale is much longer than in the explosive phase.

The \he\ found in the envelope originates from the accreted material and from
the reaction $^1$H(p,$e^-\nu)^2$H followed immediately by
$^2$H(p,$\gamma)^3$He,
which increases slightly the \he\ abundance in the initial phase of accretion.
The two major modes of \he\ destruction are through $^3$He($^3$He,2p)$^4$He
and $^3$He($^4$He,$\gamma)^7$Be. $^3$He($^4$He,$\gamma)^7$Be
is always slower than $^3$He($^3$He,2p)$^4$He, except at lower \he\ abundances.
As noted by \cite{Bof93}, the latter reaction is responsible for the
logarithmic dependence of the \be yield with respect to the initial \he
abundance above X($^3$He)$_\odot$. This means that hypothetical
higher than solar initial \he abundances, related to enriched secondary
star envelopes, do not alter dramatically the final \be yields. The results
presented in this paper are little affected by nuclear uncertainties as the
rates of the reactions of \he destruction are precisely known. 

For the ONeMg model with M=1.15 \msun, hereafter called ONe model  
since magnesium is almost absent (see \cite{Dom93} and \cite{Rit96}), 
we show the profiles of \he
and \be abundances along the envelope for different times starting
at the beginning of the accretion phase in figure \ref{f:hydro2}.
The corresponding temperatures at
the base of the envelope are 10$^7$K, 2\power{7}K, 3\power{7}K,
5\power{7}K, 10$^8$K, and the maximum temperature (2\power{8}K).
An additional model, for which a considerable expansion has already
occurred (R$_{\rm {wd}}>10^{11}$cm), is also shown.
Our results indicate that \he is destroyed down to
abundances between $10^{-6}$ and $10^{-7}$, by mass,
at the end of the accretion phase. More
specifically, these abundances
correspond to the phase during which temperatures are around $10^{8}$ K,
allowing the photodisintegration of $^8$B to prevent \be destruction
(see discussion above). Therefore, the final \be abundances
are similar to the \he ones at this critical phase. It is important to
notice that these values are much higher than those at the burning
shell ($\simeq 10^{-9}$ by mass, see figure \ref{f:hydro2}),
indicating that one--zone
models are unable to provide correct yields. The average mass fraction of \be
in the shells that will be ejected and, thus, will contribute to interstellar
medium enrichment, is around $10^{-6}$.
Since all the \li finally produced comes from the decay of \be and
as temperatures of our last model are low enough to prevent \li destruction,
the final \li yield corresponds to the addition of \be and \li mass fractions
in the ejected shells.
The final ejected mass is 1.9\power{-5}M$_{\odot}$, with a
mean abundance of \li by mass of 6.0\power{-7}. Thus, 1.1\power{-11}
M$_{\odot}$ of \li would be ejected (see table1).

Concerning the CO cases, \he destruction is less pronounced, at the same
critical phase, and this
allows the synthesis of a larger amount of \be. The corresponding
profiles of \he and \be are shown in figure \ref{f:hydro3}, for the 
M=1.15 \msun  
case. The mean mass fraction of \li is 8.2\power{-6} (the ejected mass
of \li is 1.1\power{-10}M$_{\odot}$ and the total ejected mass is
1.3\power{-5}M$_{\odot}$).
The main reason for the different nucleosynthesis in both cases is that
for CO novae the presence of $^{12}$C implies that the fast reaction
$^{12}$C(p,$\gamma)^{13}$N($\beta^{+})^{13}$C is dominant during the late
accretion phase (whereas the energy production through this reaction
is lower in ONe novae, as they almost lack from $^{12}$C). Consequently, 
the duration of the phase prior to maximum temperature is shorter in CO 
novae, preventing  an efficient \he destruction and thus leading to a 
larger final amount of \be synthesized.

As a summary of our results (see table 1, where additional cases are shown), 
overproductions of \li with respect to the solar
abundances between 100 and 2000 are obtained, depending mainly on the
chemical composition of the envelope, which is related to that of the
underlying core.  It is hard to estimate its real contribution to the \li 
enrichment in the galaxy, since theoretical models systematically produce 
ejected masses smaller than the observed ones. For instance, our models 
typically eject $\sim 10^{-5}$ \msun while the estimated mass of QU Vul 1984 
(which has been invoked as a true ''neon nova'') is around 10$^{-3}$ \msun 
(\cite{Sai92}). For an overproduction factor as large as 2000 and a total 
ejected masses of $\sim 10^{-5}$ \msun (two orders of magnitude lower than 
observed for QU Vul 1984), a nova event would produce $\sim 10^{-10}$ \msun 
of \li. If we adopt the galactic nova rate of \cite{DVa94} (20 yr$^{-1}$) and 
an age of the galaxy of $\sim 10^{10}$ years, novae should produce at least 
20 \msun of \li. This quantity is clearly smaller than the estimated present 
content of \li in the galaxy, $\sim$ 150 \msun, but, given the uncertainties 
in the ejected mass per event the contribution of 
novae to the galactic content of \li cannot be ruled out yet.
A complete analysis of \li yields by novae and their
inclusion in a model of galactic evolution is out of the scope of this
paper and will be presented elsewhere.

\section{Discussion and conclusions}

Our results confirm that nova explosions can produce significant
amounts of \li. Overproduction factors as large as 2000 are
obtained. Our results are quite different
from those obtained from one--zone models (\cite{Bof93}), as the
most important contribution to \li enrichment comes from the external
shells, where this element has been transported by convection from the
burning shell.

Comparison with the results of \cite{novae95} and \cite{Pol95} shows that 
the behavior of \be abundances can only be correctly predicted if the 
evolution of \he during the accretion phase is accurately followed. It 
is also necessary to stress that the final results strongly depend on 
the chemical composition at the onset of the explosion. If the underlying 
white dwarf is a CO one, the \li abundances are about one order of magnitude 
larger than if the white dwarf is an ONe one.

Since the decay of \be to \li emits a photon with energy
478 KeV, during a phase in which the envelope is very transparent, this
transition could be detected by the future INTEGRAL mission (with a 
sensitivity around 6\power{-6} at this energy).
The flux of the \be decay line is:

$\rm{F}(\rm{counts.s}^{-1}.cm^{-2})= 2.2\times 10^{-6} \frac{\rm {X}(^{7}Be)}
{10^{-6}} \frac{M_{ej}}{10^{-4} M_{\odot}} \frac{1}{\rm{D}^{2}(\rm{Kpc})}
e^{-t/76d}$

\noindent 
For an ejected mass of 10$^{-5}$ \msun, with an abundance by mass
of X($^{7}$Be)=8\power{-6}, the \be decay line would be detectable just after 
the outburst only for a nova closer than 0.5 Kpc. But for an ejected mass 
of 10$^{-4}$ or 10$^{-3}$ \msun, more in accordance with observations, the 
lower limit for the distance would be 1.7 or 5.4 Kpc, respectively.
This detection would provide a confirmation of the theoretical models of
novae and also ensure that \li is produced in these scenarios, encouraging
a deep search of this element in novae.

\acknowledgments

Research partially supported by the CICYT (ESP95-0091),  by the DGICYT 
(PB94-0827-C02-02), by the CIRIT
(GRQ94-8001), by the AIHF 95-335, and by the Human Capital and Mobility
Programme, (CHGE--CT92-0009) "Access to supercomputing facilities for
european researchers" established between The European Community and
CESCA/CEPBA".

\newpage
\begin{table}[htb]
\begin{center}
\caption{\li yields and ejected masses for some nova models}
\begin{tabular}{lllllll}
Comp.                                &  M$_{\rm{wd}}$(\msun)     &
$\dot{{\rm M}}$(\msyr)               &  $\overline{\rm{X}}$(\li) &
$\frac{\rm{N}(^{7}\rm{Li/H})}{N(^{7}\rm{Li/H})_{\odot}}$              &
M$^{\rm{ej}}_{\rm{tot}}$(\msun)      &  M$^{\rm{ej}}_{^{7}\rm{Li}}$(\msun)\\
\tableline
CO  &  1.0   &  2\power{-10} & 3.1\power{-6}  &  742 & 2.3\power{-5}  &
7.1\power{-11} \\
CO  &  1.15  &  2\power{-10} & 8.2\power{-6}  & 1952 & 1.3\power{-5}  &
1.1\power{-10} \\
ONe &  1.15  &  2\power{-10} & 6.0\power{-7}  & 143  & 1.9\power{-5}  &
1.1\power{-11} \\
ONe &  1.25  &  2\power{-10} & 6.5\power{-7}  & 155  & 1.8\power{-5}  &
1.2\power{-11} \\
ONe &  1.25  &  2\power{-8 } & 7.9\power{-7}  & 187  & 8.3\power{-6}  &
6.7\power{-12} \\
\end{tabular}
\end{center}
\end{table}

\newpage
\figcaption[fig1.eps]{
Profiles of \be (upper panel) and \he (lower panel) abundances along
the envelope for different times from the beginning of accretion up to 
the ejection of the envelope, for a 1.15M$_{\odot}$ ONe novae
accreting at a rate \.{M}=2\power{-10} \msyr. 
The succesive models correspond to temperatures at the
base of the envelope 2\power{7}(solid), 3\power{7} (dot), 
5\power{7} (short dash), 7\power{7} (long dash), 
10$^{8}$ (dot - short dash), 2\power{8} K (T$_{max}$) (dot - long dash) 
plus an additional case, for which 
a considerable expansion has already occurred, R$_{\rm {wd}}>10^{11}$cm 
(short dash - long dash). The upward arrow indicates 
the base of the ejected shells. 
\label{f:hydro2}
}

\figcaption[fig2.eps]{
Same as previous figure but for a CO nova.
\label{f:hydro3}
}

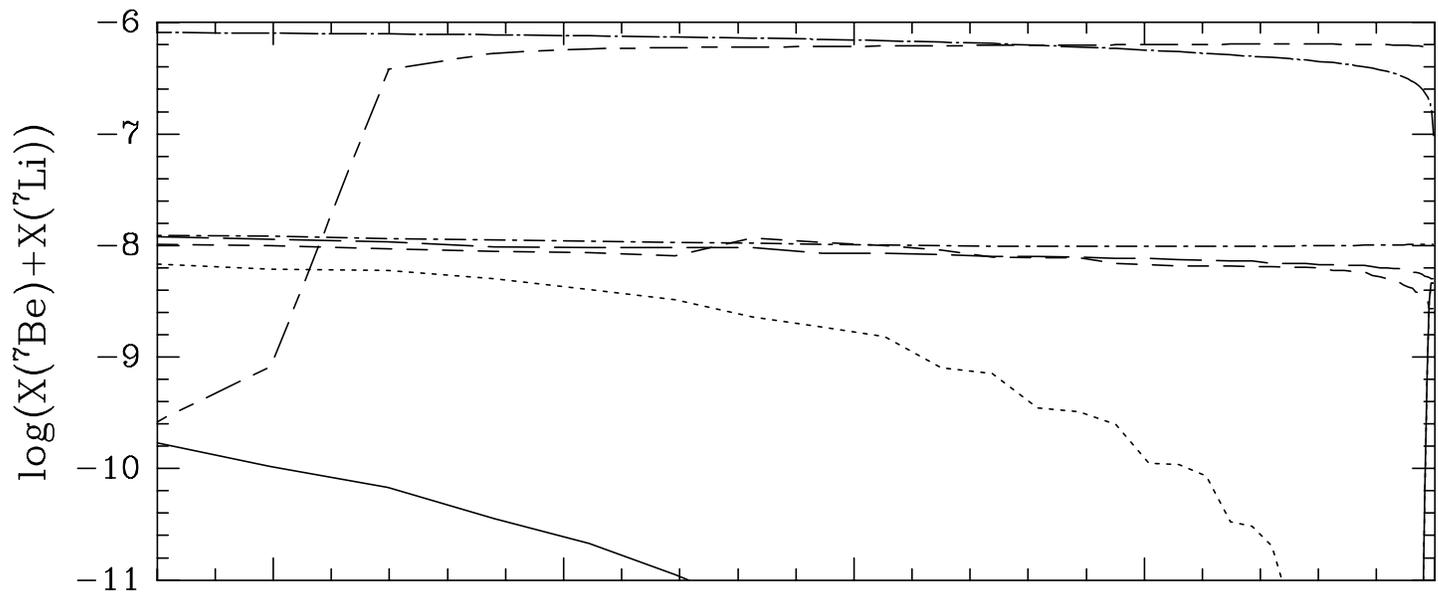
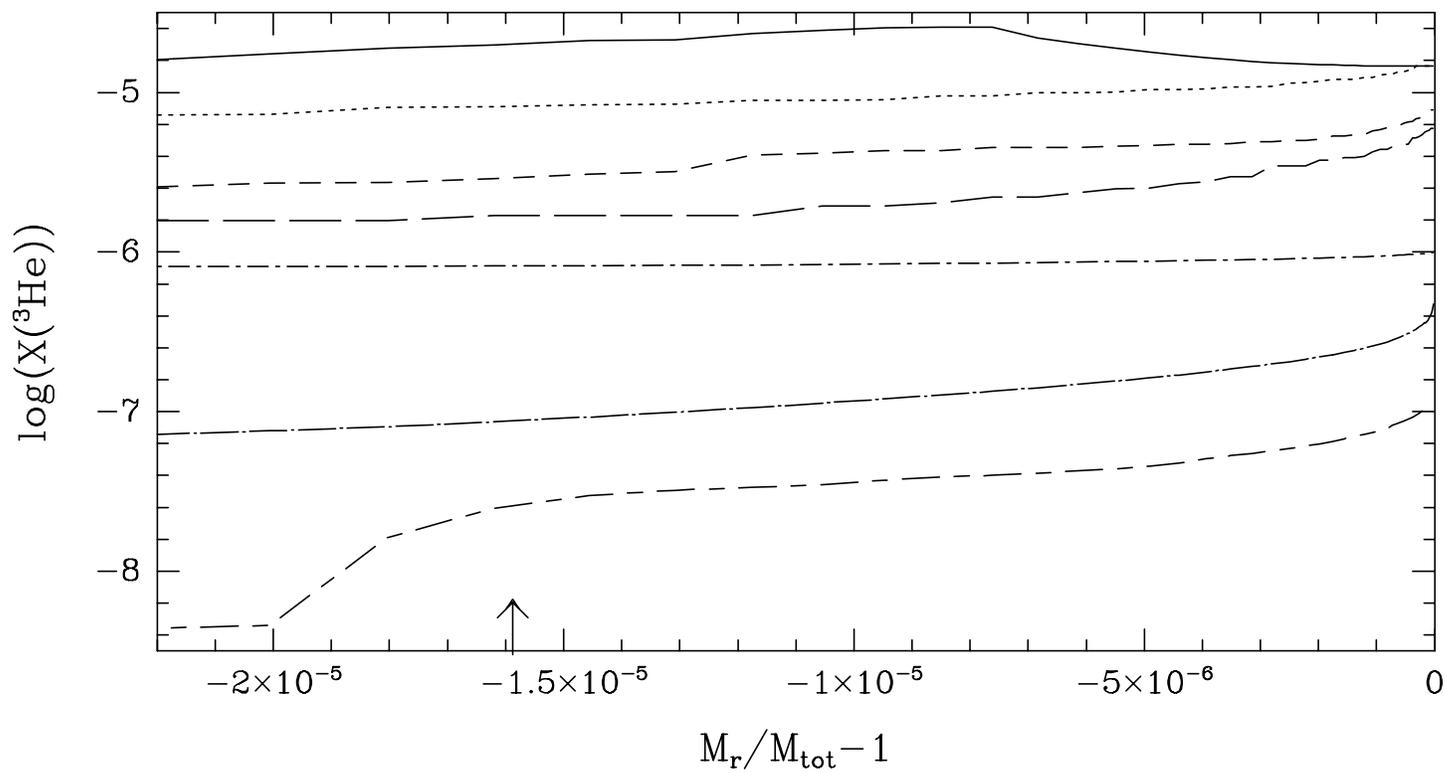
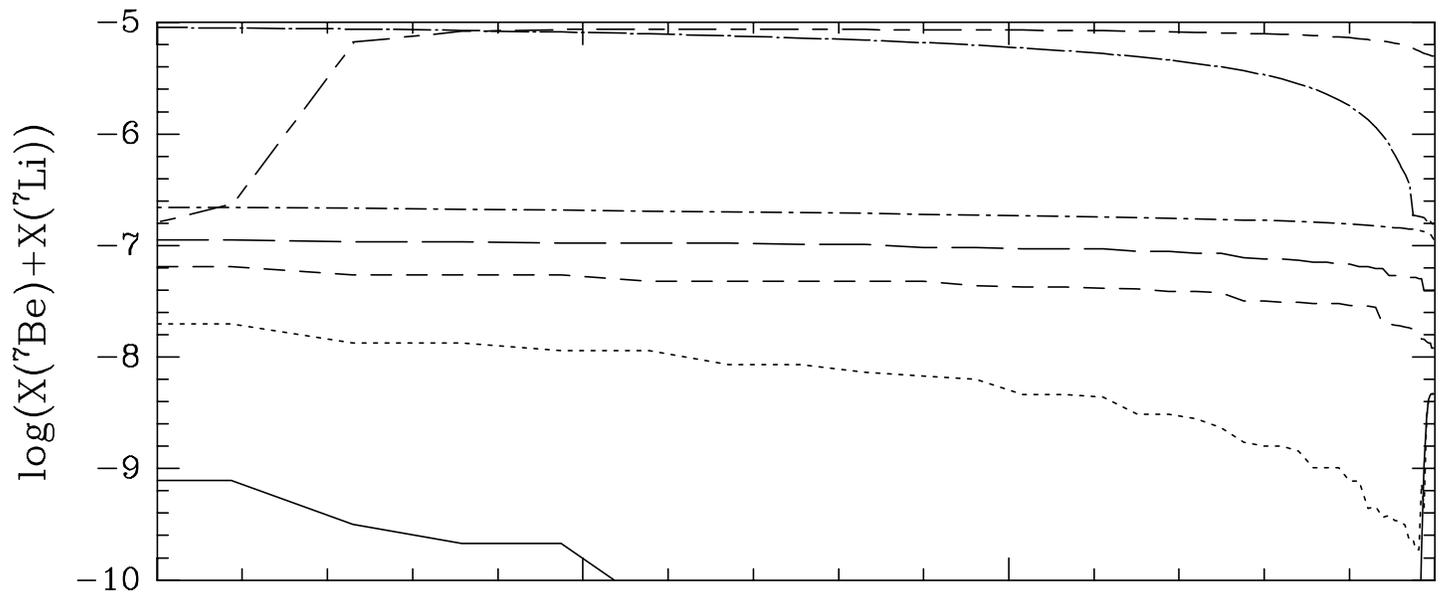
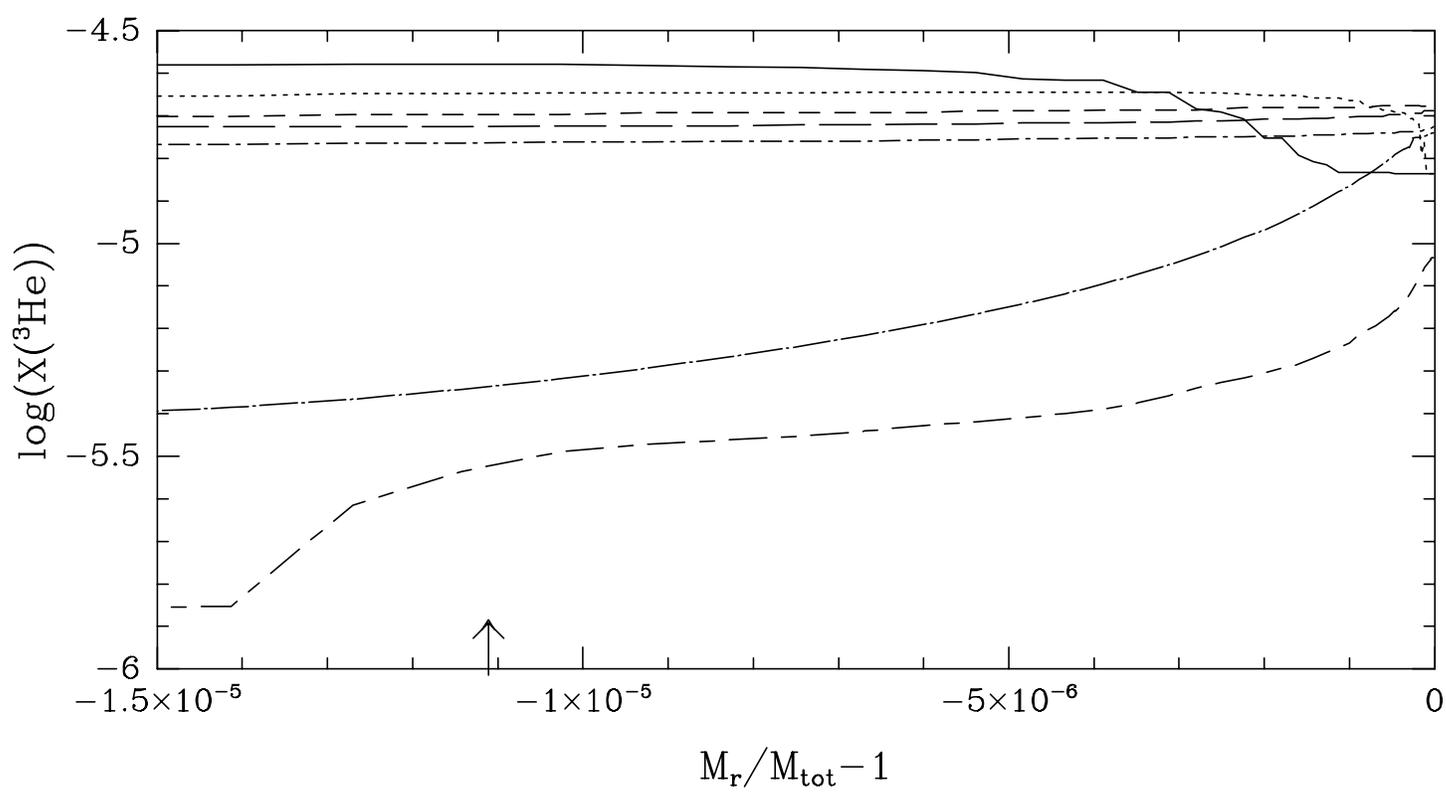
\end{document}